\documentclass[prd,a4paper,preprint,preprintnumbers,nofootinbib]{revtex4-2}

\usepackage[a4paper, hdivide={1.91cm,,1.165cm}, vdivide={1.83cm,,3.0cm}]{geometry}

\usepackage{amsmath}
\usepackage{amsfonts}
\usepackage{graphicx}
\usepackage{xspace}
\usepackage{color}
\usepackage{units}
\usepackage{slashed}
\usepackage{braket}
\usepackage[hyperfootnotes=false]{hyperref}
\hypersetup{linkcolor=red,
citecolor=green,
filecolor=cyan,
urlcolor=magenta}
\usepackage{gensymb}
\usepackage{booktabs}
\usepackage{multirow}
\usepackage{xcolor}
\usepackage{setspace}
\usepackage{physics}
\allowdisplaybreaks
\newcommand{\beq}{\begin{eqnarray}}% can be used as {equation} or  {eqnarray}
\newcommand{\eeq}{\end{eqnarray}}

\def\mcdot{\!\cdot\!}

\def\beq{\begin{equation}}
\def\eeq{\end{equation}}
\def\bea{\begin{eqnarray}}
\def\eea{\end{eqnarray}}

\begin{document}

%%%%%%%%%%%%%%%%%%%%%%%%%%%%%%%%%

\title{Analysis of FCNC  $\Xi_{QQ} \to \Lambda_Q l^+ l^-$ decay in light-cone sum rules}

\author{T.~M.~Aliev}
\email{taliev@metu.edu.tr}
\affiliation{Department of Physics, Middle East Technical University, Ankara, 06800, Turkey}

\author{S.~Bilmis}
\email{sbilmis@metu.edu.tr}
\affiliation{Department of Physics, Middle East Technical University, Ankara, 06800, Turkey}
\affiliation{TUBITAK ULAKBIM, Ankara, 06510, Turkey}

\author{M.~Savci}
\email{savci@metu.edu.tr}
\affiliation{Department of Physics, Middle East Technical University, Ankara, 06800, Turkey}

\date{\today}

\begin{abstract}
  The weak decays of doubly heavy $\Xi_{QQ}$ baryon induced by flavor changing $b \to d$ and $ c \to u$ flavor changing neutral current within the light cone sum rules are studied. The sum rules are constructed and analyzed for the corresponding transition form factors using the parallel components of the light-cone distribution amplitudes for $\Lambda_Q$ baryon. Having the results for the form factors, the corresponding branching ratios for $\Xi_{QQ} \to \Lambda_Q l^+ l^-$ decays are estimated. While the branching ratio due to $b \to d$ transition is around $10^{-9}$, in $c \to u$ transition case, it is found as around $10^{-13}$. Hence,  $\Xi_{bb} \to \Lambda_b l^+ l^-$ decay has the potential of being discovered at LHCb. However, $\Xi_{bb} \to \Lambda_b l^+ l^-$ is difficult to be measured due to the extremely small values of obtained branching ratio.
 Our results on the branching ratio are also compared with the results of the light-front approach.
\end{abstract}

%%%%%%%%%%%%%%%%%%%%%%%%%%%%%%%%%
\maketitle

\newpage
%\tableofcontents

%%%%%%%%%%%%%%%%%%%%%%%%%%%%%%%%%
%
%

\section{Introduction\label{intro}}
The quark model is a very useful tool for classifying and studying hadrons' properties. Many hadrons predicted by the quark model have already been  discovered. However, there are still many undiscovered states even though the quark model predicts them. For instance, even though doubly heavy baryons are predicted by the quark model, only $\Xi_{cc}^{++}$ baryons with mass $m_{\Xi_{cc}^{++}} = (3621.40 \pm 0.72 \pm 0.27 \pm 0.14)~\rm{MeV}$ has been observed via $\Xi_{cc}^{++} \to  \Lambda_c^+ K^- \pi^+ \pi^+$~\cite{LHCb:2017iph} at LHCb.  This state was also confirmed in the $\Xi_{cc}^{++} \to \Xi_c^+ \pi^+$~\cite{LHCb:2018pcs} and later verified in a series of experiments~\cite{LHCb:2018zpl,LHCb:2019epo,LHCb:2019ybf,LHCb:2019gqy}.

Doubly heavy baryons are one of the subjects studied intensively in many works (see~\cite{Xing:2018lre,Shi:2019fph,Aliyev:2022rrf} and references therein). The process induced by the flavor changing neutral current (FCNC) at quark level induced by $b \to d/s$ and $c \to u$ transitions occur at loop level in the standard model; hence, their decay widths are expected to be small.

%%%%%%%%%%%%%%%%%%%%%%%%%%%%%%%%%%%%%%%%%%%%%%%%%%%%%%%%%%%%%%%%%%%%%%%%%5

In the diquark-quark picture, the quantum numbers of possible doubly heavy baryons with spin-1/2 and spin-3/2 are presented in Table~\ref{tab:1}.

\begin{table*}[hbt]
  \centering
  \renewcommand{\arraystretch}{1.2}
  \setlength{\tabcolsep}{7pt}
  \begin{tabular}{ccccc}
    \toprule
     Doubly Heavy Baryon    & $J^P$             & Doubly Heavy Baryon         & $J^P$             \\
    \midrule
    $\Xi_{QQ}$              & $\frac{1}{2}^{+}$ & $\Omega_{QQ}$               & $\frac{1}{2}^{+}$ \\
    $\Xi_{QQ}^{*}$          & $\frac{3}{2}^{+}$ & $\Omega_{QQ}^{*}$           & $\frac{3}{2}^{+}$ \\
    $\Xi_{QQ^\prime}$        & $\frac{1}{2}^{+}$ & $\Omega_{QQ^\prime}$        & $\frac{1}{2}^{+}$ \\
    $\Xi^\prime_{QQ^\prime}$  & $\frac{1}{2}^{+}$ & $\Omega^\prime_{QQ^\prime}$ & $\frac{1}{2}^{+}$ \\
    $\Xi_{QQ^\prime}^{*}$    & $\frac{3}{2}^{+}$ & $\Omega_{QQ^\prime}^{*}$    & $\frac{3}{2}^{+}$ \\
    \bottomrule
  \end{tabular}
  \caption{Quantum numbers of ground state doubly heavy baryons}
  \label{tab:1}
\end{table*}

%%%%%%%%%%%%%%%%%%%%%%%%%%%%%%%%%%%%%%%%%%%%%%%%%%%%%%%%%%%%%%%%%%%%%%%%%5
The FCNC induced by doubly heavy baryons processes deserves much theoretical and experimental attention.
The weak decays of doubly heavy baryons induced by FCNC transitions are a significant class of decays for studying the properties of such baryons. Analyzing these channels can give useful information about the helicity structure of effective Hamiltonian and Cabibbo-Kobayashi-Maskawa (CKM) matrix elements and also look for new physics effects.

While the semileptonic decays induced by charged current occur at tree level, the decays induced by flavor-changing neutral current take place at loop level. The leptonic part of these decays is well known, and all the QCD dynamics are encoded in the hadronic matrix elements induced by the weak current. These matrix elements are parameterized in terms of the form factors that play a crucial role in analyzing the semileptonic and non-leptonic decays. Therefore determining these  form factors constitutes a central problem in studying such a class of decays.

The form factors belong to the non-perturbative domain of QCD, and some non-perturbative methods are needed for their determination. QCD sum rule is one of the powerful methods that take into account the non-perturbative effects~\cite{Shifman:1978bx}. One of the modifications of the traditional QCD sum rules method is the light cone version~\cite{Braun:1997kw} where operator product expansion(OPE) is performed over the twist of operators rather than the dimensions of operators.

The form factors induced by charged currents for $\Xi_{QQ^\prime} \to \Lambda_Q^\prime$ are estimated within light-cone sum rules~\cite{Hu:2019bqj}, in QCD sum rules methods~,\cite{Hu:2019bqj} and in light-front formalism~\cite{Shi:2019hbf}. A similar analysis for $\Xi_{QQ^\prime} \to \Sigma_Q^\prime$ transition within the light-front quark model and light cone QCD sum rules are investigated in~\cite{Zhao:2018mrg,Wang:2017mqp}, and \cite{Shi:2019fph} respectively.

FCNC processes of doubly heavy baryons in the framework of the light-front approaches are comprehensively studied in~\cite{Xing:2018lre}. In the present work, we apply light cone QCD sum rules to study doubly heavy baryon decays induced by FCNC transition. The paper is organized as follows. In section~\ref{sec:2}, we present the effective Hamiltonian responsible for $b \to s/d l^+ l^-$ and $c \to u l^+ l^-$ transition. Then, the transition form factors are derived. The numerical results for the form factors are presented in Section~\ref{sec:3}. The final section contains the summary and comparison of our findings with  the predictions of other approaches.

%%%%%%%%%%%%%%%%%%%%%%%%%%%%%%%%%%%%%%%%%%%%%%%%%%%%%%%%%%%%% 

\section{Light cone sum rules for the transition form factors}
\label{sec:2}
The effective Hamiltonian responsible for $b \to q~\text{(q = s or d)}$ is given by
\begin{equation}
  \label{eq:1}
  H_{eff} = \frac{G_F}{\sqrt{2}} V_{tb} V_{tq}^* \sum_{i=1}^{10} C_i(\mu) \mathcal{O}_i(\mu)
\end{equation}
where $G_F$ is the Fermi coupling constant, $V_{tb}$ and $V_{tq}$ are the elements of CKM matrix elements. $\mathcal{O}_i$ corresponds to the local operators and $C_i$ are their Wilson coefficients. Their explicit forms $\mathcal{O}_i$ and $C_i$ can be obtained in~\cite{Buchalla:1995vs}. The transition amplitude for $\Xi_{bb} \to \Lambda_{b} l^+ l^-$ is given 
\begin{equation}
  \label{eq:2}
  \begin{split}
    \mathcal{M}(\Xi_{bb} \to \Lambda_b l^+ l^-) &= \frac{G_f \alpha_{em}}{2 \sqrt{2} \pi} V_{tb}V_{td}^* \times \bigg\{  \bigg[ C_9^{\text{eff}} \bra{\Lambda_b}  \bar{d} \gamma_\mu (1-\gamma_5) b  \ket{\Xi_{bb}}   \\
    & - 2 m_b C_{7}^{\text{eff}} \bra{\Lambda_b}  \bar{d} i \sigma_{\mu \nu} \frac{q^\nu}{q^2}(1+\gamma_5) b   \ket{\Xi_{bb}} \bigg] \bar{l} \gamma_\mu l \\
    &+ C_{10} \bra{\Lambda_b}  \bar{d} \gamma_\mu (1- \gamma_5) b  \ket{\Xi_{bb}} \bar{l} \gamma_\mu \gamma_5 l \bigg\}
\end{split}
\end{equation}
The transition amplitude for $\Xi_{cc} \to \Lambda_c l^+ l^-$ induced by $c \to u $ transition can be obtained form Eq.~\ref{eq:2} by replacing $V_{tb}V_{td}^*$ by $\sum_{i} V_{ci} V_{ui}^*$. Obviously, the numerical values of corresponding Wilson coefficients will be different than for $ b \to d$ case and the values of $C_9$, $C_7$ and $C_{10}$ for $c \to u$ transition can be found in~\cite{Burdman:2001tf,Burdman:1995te,Fajfer:1999dq}.

The matrix element $ \bra{\Lambda_b(p)}  \bar{d} \gamma_\mu (1-\gamma_5) b | \ket{ \Xi_{bb}(p+q)}$ where initial and final baryons are spin-1/2 states is parameterized in terms of the six form factors as follows.
\begin{equation}
  \label{eq:5}
  \begin{split}
    \bra{B_{Q^\prime}(p)}   \bar{q} \gamma_\mu (1-\gamma_5) Q  \ket{ \Xi_{bb}(p+q)} & =
    \bar{u}(p) \bigg[ \gamma_\mu f_1(q^2) + i \sigma_{\mu \nu} \frac{q^\nu}{M}f_2 + \frac{q_\mu}{M}f_3 \\
    &- \big(\gamma_\mu g_1(q^2) + i \sigma_{\mu \nu} \frac{q^\nu}{M} g_2 + \frac{q_\mu}{M}g_3 \big) \gamma_5 \bigg] u(p+q)
  \end{split}
\end{equation}
where $q_\mu = p^\prime - p$ and M is the mass for doubly heavy baryon.

The matrix element $\bra{\Lambda_b(p)} | \bar{q} i \sigma_{\mu \nu} q^\nu (1+\gamma_5) Q | \ket{ \Xi_{bb}(p)}$ is expressed in terms of the four form factors in the following way
\begin{equation}
  \label{eq:6}
  \begin{split}
    \bra{B_{Q^\prime}(p)}   \bar{q} i \sigma_{\mu \nu} q^{\nu} (1+\gamma_5) Q   \ket{ \Xi_{bb}(p^\prime)} = \bar{u}(p) \bigg[ 
    & \frac{f_1^T}{M} (\gamma_\mu q^2 - q_\mu \slashed{q}) + i f_2^T \sigma_{\mu \nu} q^\nu  \\
    &+ \frac{g_1^{T}}{M} (\gamma_\mu q^2 - q_\mu \slashed{q}) \gamma_5 + i g_2^T \sigma_{\mu \nu} q^\nu \gamma_5 \bigg]
  \end{split}
\end{equation}
Before starting our analysis, we would like to note that the form factors $f_i$, $g_i$ entering to Eq.\eqref{eq:5} induced by $\bar{q} \gamma_\mu (1 - \gamma_5) Q$ current within the same framework is studied in~\cite{Shi:2019fph} and therefore, we will concentrate only on the calculation of $f_i^T$ and $g_i^T$ form factors. The calculations of the form factors are the main issue in analyzing semileptonic decays of baryons.
In the present work, we employ the light-cone sum rules method (LCSR) to calculate the $f_i^T$ and $g_i^T$ form factors. For this goal, we consider the following correlation function 
\begin{equation}
  \label{eq:35}
  \begin{split}
    \Pi_\mu (p,q) &= i \int d^4x e^{iqx} \bra{B_Q^\prime (p)}  T \big\{ \bar{q} i \sigma_{\mu \alpha} q^{\alpha}(1+ \gamma_5) Q \bar{J}_{Q Q^\prime}(0) \big\} \ket{0} \\ 
  \end{split}
\end{equation}
Here, $J_\mu^{V-A}$, is the transition current, and  $J_{QQ^\prime}$ is the interpolating current for the spin-$1/2$ doubly heavy baryon.

For $Q= Q^\prime = b \text{ or } c$, the interpolating currents of spin-$1/2$ are
\begin{equation}
  \label{eq:36}
  \begin{split}
    J_{QQ} = \epsilon^{abc} ({Q^a}^T C \gamma^\alpha Q^b) \gamma_\alpha \gamma_5 q^c
  \end{split}
\end{equation}
where $q = u$ or $d$ for $\Xi_{QQ}$ baryon.
To construct the LCSR for the relevant form factors, the correlation functions should be calculated both from the hadronic and QCD sides. Then, the results of these two representations are matched according to quark-hadron duality ansatz.

The representation of the correlation function from the hadronic part is obtained by inserting a complete set of baryon states and isolating the ground state of $\Xi_{bb}$ baryon. 
As a result, for the hadronic part of the correlation function $\Pi_\mu$, we get
\begin{equation}
  \label{eq:33}
  \begin{split}
\Pi_\mu^{II} &= {\lambda\over M^2-(p+q)^2} \bar{u}(v)
\Big\{ \Big[\Big( {m \over M} - 1 \Big) f_1^T + f_2^T \Big] \slashed{q} q_\mu
+ 2 m f_2^T \slashed{q} v_\mu  \\
&+ \Big[\Big( {m \over M} + 1 \Big) g_1^T + g_2^T \Big] \slashed{q}
\gamma_5 q_\mu - 2 m g_2^T \slashed{q} \gamma_5 v_\mu  +
\mbox{other structures}   \Big\}
  \end{split}
\end{equation}
where $v$ is the velocity of the heavy baryon, $\lambda$, and $M$ are the residue and mass of doubly heavy baryons, and $m$ corresponds to the mass of a single heavy baryon.
To derive this equation, we used the definition of the decay constant
\begin{equation}
  \label{eq:14}
  \bra{B(p+q)_{Q^\prime Q}} \bar{J}_{BB^\prime} \ket{0} = \lambda \bar{u}(p+q)
\end{equation}

Now let us  turn our attention to the calculation of the correlation function from the QCD side. To achieve this, we used the operator product expansion (OPE) carried over twists of non-local  operators and involves the single heavy baryon distribution amplitudes. Before presenting the details of the calculation of the correlation function from the QCD side, a few words about DA's of $\Lambda_Q$ baryon is in order. 
The light cone distribution amplitudes for single-heavy baryons are obtained in~\cite{Ball:2008fw,Ali:2012pn} by using the sum rules method at the heavy quark mass limit. In this study, we used the DA's of $\Lambda_Q$ obtained in~\cite{Ball:2008fw}. The DA's of anti-triplet $\Lambda_Q$ baryon are defined with the help of four matrix elements of non-local operators:
\begin{equation}
 \label{eq:22}
  \begin{split}
  \frac{1}{v_{+}} \langle 0 | \left[q_{1}\left(t_{1} \right) \mathcal{C} \gamma_{5} \slashed{n} \slashed{q}_{2} \left(t_{2} \right) \right]   Q_{\gamma} |  H_{b}^{j=0} \rangle &=\psi_2 \left(t_{1} , t_{2}\right) f_{H_{b}^{j=0}}^{(1)} u_{\gamma} \\
\frac{i}{2}\langle 0|   \left[q_{1} \left( t_{1} \right) \mathcal{C} \gamma_{5} \sigma_{\bar{n} n} q_{2} \left(t_{2}\right) \right] Q_{\gamma} | H_{b}^{j=0}  \rangle &= \psi_3^{\sigma} \left(t_{1}, t_{2}\right) f_{H_{b}^{j=0}}^{(2)} u_{\gamma}  \\
   \langle 0 | \left[q_{1}\left(t_{1}\right) \mathcal{C} \gamma_{5} q_{2}\left(t_{2}\right)\right] Q_{\gamma} | H_{b}^{j=0}\rangle  &=\psi_3^{s} \left(t_{1}, t_{2}\right) f_{H_{b}^{j=0}}^{(2)} u_{\gamma} \\
   v_{+}\langle 0 | \left[q_{1}\left(t_{1}\right) \mathcal{C} \gamma_{5} \slashed{n} q_{2}\left(t_{2}\right)\right] Q_{\gamma} | H_{b}^{j=0} \rangle &=\psi_4 \left(t_{1} , t_{2}\right) f_{H_{b}^{j=0}}^{(1)} u_{\gamma}
 \end{split}
\end{equation}
where $n$ and $\bar{n}$ are two light-cone vectors, $\bar{v}^\mu = \frac{1}{2} \big( \frac{n^\mu}{v_{+} - v_{+} \bar{n}^\mu} \big)$, $t_i$ are the differences between the $i$th light quark and origin along the direction of $n$. The space coordinates of light quarks is $t_i n^\mu$. The four velocity of single baryon is defined as $v^{\mu} = \frac{1}{2} \big(\frac{n^\mu}{v_{+}} + v_{+} \bar{n}^\mu \big)$. In our analysis, we will work in the rest frame of single heavy baryon, i.e. $v_{+} = 1$. In addition  $\psi_2, \psi_3^{\sigma}, \psi_3^s$, and $\psi_4$ denote the DA's with twists 2, 3, and 4 respectively.. In present work, to determine the transition form factors we only use the parallel DA's for $\Lambda_Q$ baryon.

The following matrix element $\epsilon^{a b c}\left\langle \Lambda_{Q}(v)\left|\bar{q}_{1 \alpha}^{a}\left(t_{1}\right) \bar{q}_{2 \beta}^{b}\left(t_{2}\right) \bar{Q}_{\gamma}^{c}(0)\right| 0\right\rangle$ in the calculation of the correlation function. This matrix element is determined in terms of the heavy baryon distribution amplitudes as follows
\begin{equation}
\begin{aligned}
\epsilon^{a b c}\left\langle \Lambda_{Q}(v)\left|\bar{q}_{1 \alpha}^{a}\left(t_{1}\right) \bar{q}_{2 \beta}^{b}\left(t_{2}\right) \bar{Q}_{\gamma}^{c}(0)\right| 0\right\rangle &= \frac{1}{8}   \psi_{2}^{*}\left(t_{1}, t_{2}\right) f^{(1)} \bar{u}_{\gamma}\left(C^{-1} \gamma_{5} \bar{\slashed{n}} \right)_{\alpha \beta} \\
&-\frac{1}{8} \psi_{3 \sigma}^{*}\left(t_{1}, t_{2}\right) f^{(2)} \bar{u}_{\gamma}\left(C^{-1} \gamma_{5} i \sigma^{\mu \nu}\right)_{\alpha \beta} \bar{n}_{\mu} n_{\nu} \\
&+\frac{1}{4} \psi_{3 s}^{*}\left(t_{1}, t_{2}\right) f^{(2)} \bar{u}_{\gamma}\left(C^{-1} \gamma_{5}\right)_{\alpha \beta} \\
&+\frac{1}{8  } \psi_{4}^{*}\left(t_{1}, t_{2}\right) f^{(1)} \bar{u}_{\gamma}\left(C^{-1} \gamma_{5} \slashed{n} \pi\right)_{\alpha \beta}
\end{aligned}
\end{equation}
The DA's $\psi(t_1,t_2)$ is defined as
\begin{equation}
  \label{eq:15}
  \psi(t_1,t_2) = \int_0^\infty d\omega \omega \int_0^1 du e^{-i \bar{u} \omega v (x_2 - x_1)} e^{-i w v x_1} \psi(\omega,u)
\end{equation}
where $\omega$ is the total diquark momentum. 
Using the expressions of the interpolating current for $\Xi_{QQ}$ baryon as well as the transition current, the correlation function can be written as
\begin{equation}
  \label{eq:16}
  \begin{split}
    \Pi_{\mu}(p,q) &= - \frac{i}{4} \int d^4x \int_0^\infty d\omega \omega \int_0^1 du e^{i(q+\bar{u} \omega v) x}  \times 
    \bigg\{ \bar{u}  \sum_{i=1}^4 a_i (\gamma_\nu C)^T S_Q^T(x) \mathcal{T}_\mu^{T} (C^{-1} \gamma_5 \Gamma_i)^T \gamma_\nu \gamma_5 \bigg\}
  \end{split}
\end{equation}
\begin{equation}
  \label{eq:17}
  \begin{aligned}
 & \mathcal{T}_{\mu } = i \sigma_{\mu \rho} q^\rho (1 + \gamma_5), \\
    a_1 &= f^{(1)} \psi_2                & \Gamma_1 &= \bar{\slashed{n}}, \\
    a_2 &= - f^{(2)} \psi_{3\sigma}                    & \Gamma_2 &= i \sigma_{\alpha \beta} \bar{n}^\alpha n^{\beta}, \\
    a_3 &= 2 f^{(2)} \psi_{3s}                          & \Gamma_3 &= 1, \\
    a_4 &= -f^{(1)} \psi_{4}             & \Gamma_4 &= \slashed{n},
  \end{aligned}
\end{equation}
and $S_{Q}(x)$ is the free heavy quark propagator. $\bar{v}$ as well as the light-cone vectors $n$, $\bar{n}$ are defined as
\begin{equation}
  \label{eq:18}
  \begin{split}
    n_{\mu} &= \frac{1}{v x} v_{\mu} \\
    \bar{n}_{\mu} &= 2 v_{\mu} - n_\mu \\
    \bar{v}_\mu &= n_\mu - v_\mu
  \end{split}
\end{equation}
Taking Fourier transformation of heavy quark propagator, we get
\begin{equation}
  \label{eq:19}
  \begin{split}
  \Pi_{\mu}(p,q) &= \frac{1}{4} \int d^4x \int d\omega \omega \int du \int \frac{d^4k}{(2\pi)^4} e^{i(q + \bar{u} w v -k)x } \times \\ \Bigg[ 
  & \bar{u} \sum_{i=1}^{4} a_i \gamma_\nu \frac{\slashed{k} - m_Q}{k^2 - m_Q^2} C \mathcal{T}_{\mu}^{T} \Gamma_i^{T} \gamma_5^{T} C^{-1} \gamma_\nu \gamma_5 \Bigg] 
  \end{split}
\end{equation}
After performing integration over $x$, one can obtain the explicit expression for the correlation function from the QCD side as follows.
\begin{equation}
  \label{eq:34}
  \begin{split}
    \Pi_\mu^{QCD} [(p+q)^2,q^2] =
\int du \int dw {\sum_{n=1}^{2} }  \Bigg\{&  {\rho_n^{(1)}(u,w) \over (\Delta - m_Q^2)^n}  \slashed{q} q_\mu 
+  {\rho_n^{(2)}(u,w) \over (\Delta - m_Q^2)^n} \slashed{q} v_\mu \\
&+
{\rho_n^{(3)}(u,w) \over (\Delta - m_Q^2)^n} \slashed{q} \gamma_5 q_\mu 
+ {\rho_n^{(4)}(u,w) \over (\Delta - m_Q^2)^n} \slashed{q} \gamma_5v_\mu  \\
&+ \text{other structures}~,
\Bigg\}
  \end{split}
\end{equation}
where $\Delta = (q + \bar{u} w v)^2$  and $\bar{u} = 1-u$. 
The explicit expressions of $\rho_n^{(i)}(u,w)$ are presented in Appendix~\ref{sec:appa}.
Choosing the coefficients of the structures $\slashed{q} q_{\mu}$, $\slashed{q} v_\mu$, $\slashed{q} \gamma_5 q_\mu$ and $ \slashed{q} \gamma_5 v_\mu$ in both sides of the correlation function and performing Borel transformation on variable $-(p+q)^2$ we get the following sum rules for the form factors $f_1^T$, $f_2^T$, $g_1^T$ and $g_2^T$ :
\begin{equation}
  \label{eq:21}
  \begin{split}
   \lambda  \Big[ f_1^T \Big( \frac{m}{M} - 1\Big) + f_2^T \Big] e^{-\frac{M^2}{M_B^2}}  &= \Pi_1^B  \\
 2\lambda m f_2^T e^{-\frac{M^2}{M_B^2}} &= \Pi_2^B  \\
   - \lambda \Big[ g_1^T \Big( \frac{m}{M} + 1 \Big) + g_2^T \Big] e^{-\frac{M^2}{M_B^2}}  &=   \Pi_3^B \\
 - 2 m \lambda  g_2^T e^{-\frac{M^2}{M_B^2}} &=   \Pi_4^B 
  \end{split}
\end{equation}
where $\Pi_i^B$ is the Borel transformed coefficients of structures mentioned above from the QCD side, and $m_B$ is the Borel mass parameter. The Borel transformation is performed with the help of the master formula
\begin{equation}
\begin{aligned}
\int dw \frac{\rho_n(u,w)}{(\Delta - m_Q^2)^n} =& \sum_{n=1}^{\infty}\left\{(-1)^{n} \int_{0}^{w_{0}} d w e^{\left(-s\left(w, q^{2}\right) \right) / M_B^{2}} \frac{1}{(n-1) !\left(M_B^{2}\right)^{n-1}} \frac{\rho_n(u, w)}{(\bar{u} w /m)^n} \right.\\
&\left.-\left[\frac{(-1)^{n-1}}{(n-1) !} e^{\left(-s\left(w, q^{2}\right) \right) / M_B^{2}} \sum_{j=1}^{n-1} \frac{1}{\left(M_B^{2}\right)^{n-j-1}} \frac{1}{s^{\prime}}\left(\frac{\mathrm{d}}{\mathrm{d} w} \frac{1}{s^{\prime}}\right)^{j-1} \frac{\rho_n(u, w)}{(\bar{u} w /m)^n} \right]_{w=w_{0}}\right\},
\end{aligned}
\end{equation}
where $s(u,\omega) = \frac{m_Q^2 - \bar{u} \omega (\bar{u} \omega - m) - (1 - \frac{\bar{u} \omega}{m}) q^2}{ \frac{ \bar{u} \omega}{m}}$ , and $s^\prime = \frac{ds}{d\omega}$, and $\omega_0$ are the solutions of the $s_{th} = s$ and $s = 4 m_Q^2$ equations respectively.

\section{Numerical Analysis}
\label{sec:3}
In this section, first, we perform a numerical analysis of the transition form factors. Then using the results for the form factors, we estimate the decay width of $\Xi_{QQ} \to \Lambda_Q l^+ l^-$ transitions. The main parameters entering the sum rules are the mass of the heavy quarks, the mass of the doubly heavy baryon and its residue, and the decay constants $f^{(1)}$ and $f^{(2)}$ of $\Lambda_Q$ baryon. For the masses of $c$ and $b$ quarks, we used their values in $\overline{MS}$ scheme, i.e. $\overline{m}_c(\overline{m}_c)$ = $1.275 \pm 0.025~GeV$ and $\overline{m}_b(\overline{m}_b)$ = $4.18 \pm 0.03~GeV$ ~\cite{PhysRevD.98.030001}, respectively. Moreover, the mass and decay constant $\lambda$ of doubly heavy baryons are~\cite{Shi:2019fph}:
\begin{align}
  \label{eq:3}
M_{\Xi_{cc}} &= 3.621~\rm{GeV} &      \lambda_{\Xi_{cc}} &= 0.109 \pm 0.020 \nonumber \\  
M_{\Xi_{bb}} &= 10.143~\rm{GeV} &     \lambda_{\Xi_{bb}} &= 0.190 \pm 0.052 \nonumber~. 
\end{align}
The mass values of the $\Lambda_c$ and $\Lambda_b$ baryons are taken as~\cite{PhysRevD.98.030001} $m_{\Lambda_c} = 2.286~\rm{GeV}$ and $m_{\Lambda_b} = 5.620~\rm{GeV}$. For the decay constants of $\Lambda_c$ and $\Lambda_b$ baryons we used, $f^{(1)} = f^{(2)} = (2.8 \pm 0.2) \times 10^{-2}~\rm{GeV^3}$~\cite{Groote:1996em}.

The main non-perturbative input parameters of any light-cone sum rules are the light-cone distribution amplitudes (DA's) of corresponding baryons. Although the DA's of $\Lambda_c$ have not been known yet, thanks to the heavy quark limit, they are supposed to have the same form as $\Lambda_b$. Hence, we can use the same form of DA's presented in~\cite{Ali:2012pn,Ball:2008fw} for both $\Lambda_b$ and $\Lambda_c$ baryons.

\begin{equation}
  \label{eq:7}
  \begin{split}
    \psi_{2}(\omega, u) &= \frac{15}{2} A^{-1} \omega^{2} \bar{u} u \int_{\omega / 2}^{s_{0}} d s e^{-s / y}(s-\omega / 2) \\
    \psi_{4}(\omega, u) &= 5 A^{-1} \int_{\omega / 2}^{s_{0}} d s e^{-s / y}(s-\omega / 2)^{3}, \\
    \psi_{3 s}(\omega, u) &= \frac{15}{4} A^{-1} \omega \int_{\omega / 2}^{s_{0}} d s e^{-s / y}(s-\omega / 2)^{2} \\
\psi_{3 \sigma}(\omega, u) &= \frac{15}{4} A^{-1} \omega(2 u-1) \int_{\omega / 2}^{s_{0}} d s e^{-s / y}(s-\omega / 2)^{2}
  \end{split}
\end{equation}
where 
\begin{equation}
  \label{eq:8}
  A=\int_{0}^{s_{0}} d s s^{5} e^{-s / y}
\end{equation}
and $y$ and $s_{0}$ are the Borel parameter and the continuum threshold introduced by QCD sum rules in~\cite{Ball:2008fw,Ali:2012pn}. While $y$ is varied in the interval $0.4~\mathrm{GeV} < y < 0.8~\mathrm{GeV}$ a fixed value $s_{0} = 1.2~\mathrm{GeV}$ is used. Note that the DA's are only non-vanishing in the region $0<\omega < 2 s_{0}$~\cite{Ali:2012pn}.
%%%%%%%%%%%%%%%%%%%%%%%%%

In addition to these input parameters, the sum rules for the form factors contain two auxiliary parameters, continuum thresholds $s_{th}$ and the Borel mass parameters. The choice of the working regions of $M_{B}^2$ regions is based on the standard criteria; namely, both power corrections and continuum contributions should be sufficiently suppressed.

The upper limit of $M_{B}^2$ is determined from the condition that the contributions of continuum and excited states are less than $30 \%$ of the total LCSR result, i.e., $\frac{\Pi(M_{B}^2,s_{th})}{\Pi(M_{B}^2,\infty)} \lesssim 0.3$. The lower limit of $M_B^2$ is obtained by requiring the convergence of operator product expansion near the light cone for the correlation in the deep Euclidean domain, i.e., the contributions of the higher twist amplitudes constitute maximum $10 \%$ of the leading twist contribution,  $\frac{\Pi(M_{B}^2, \text{twist-4 DA's})}{\Pi(M_{B}^2, \text{twist-2 DA's})} \lesssim 0.1$. Here $\Pi(M_{B}^2, \text{twist-4 DA's})$ and $\Pi(M_{B}^2, \text{twist-2 DA's)}$ correspond to the higher twist-4 and leading twist-2 DA's contributions, respectively. Both conditions are simultaneously satisfied in the regions depicted in Table~\ref{tab:3}. The value of $s_{th}$ is chosen in such a way that the differentiated mass sum rules reproduce $10\%$ accuracy of the corresponding baryon. Considering these criteria, we obtain the working region for $s_{th}$ that is presented in Table~\ref{tab:3}.

\begin{figure}[t]
\includegraphics[width=0.46\textwidth]{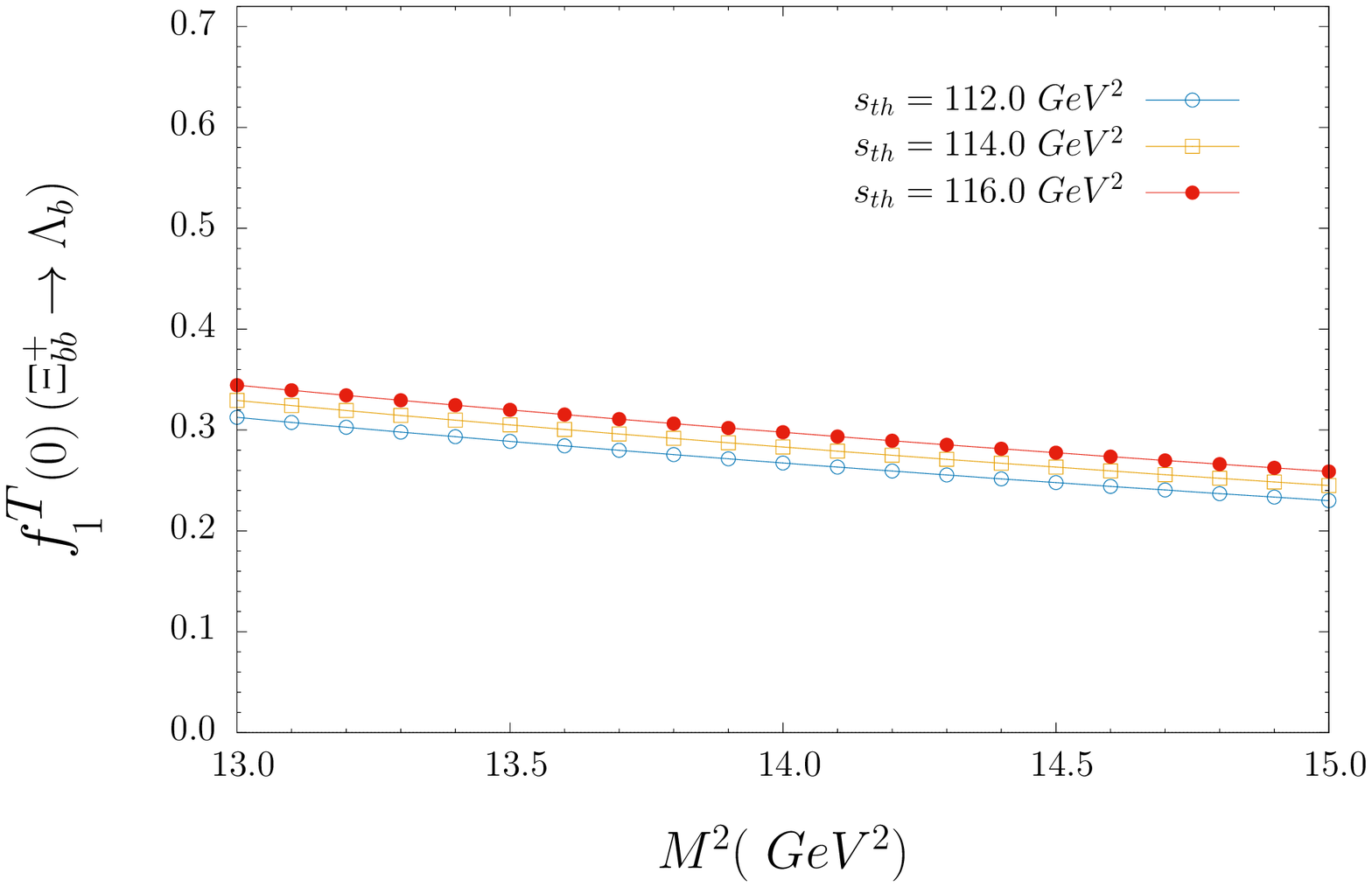}
\includegraphics[width=0.46\textwidth]{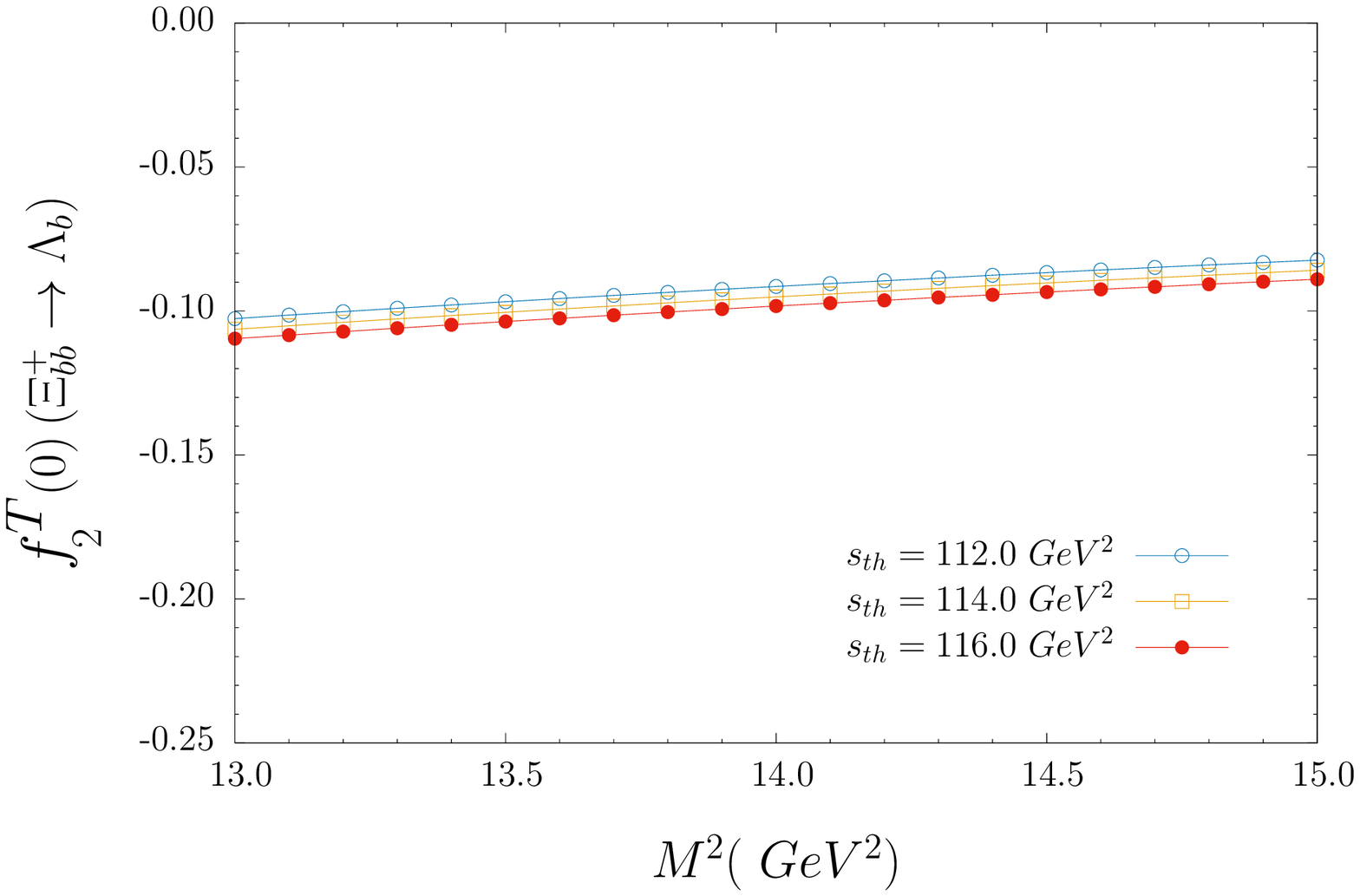} \\
\includegraphics[width=0.46\textwidth]{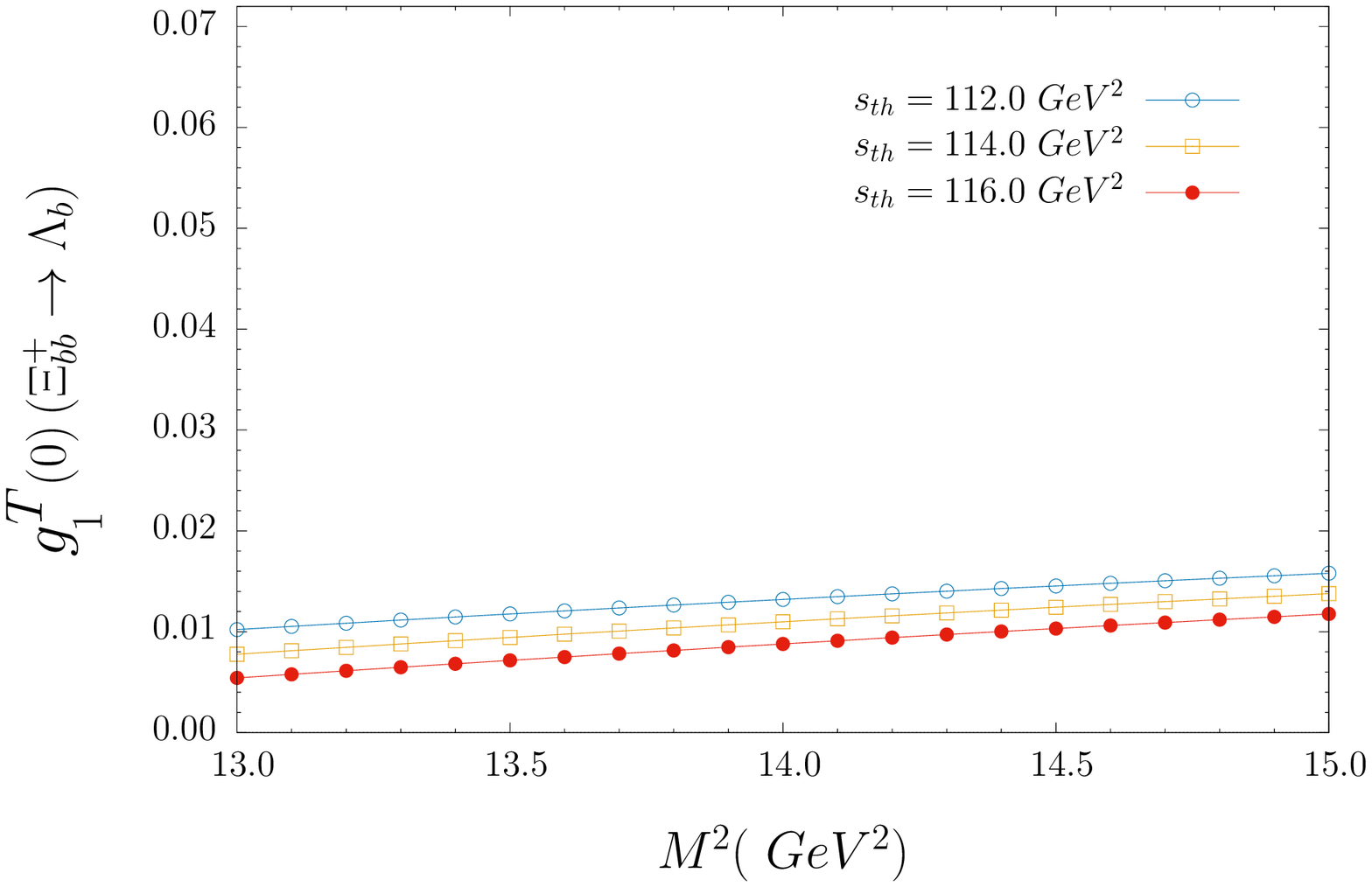}
\includegraphics[width=0.46\textwidth]{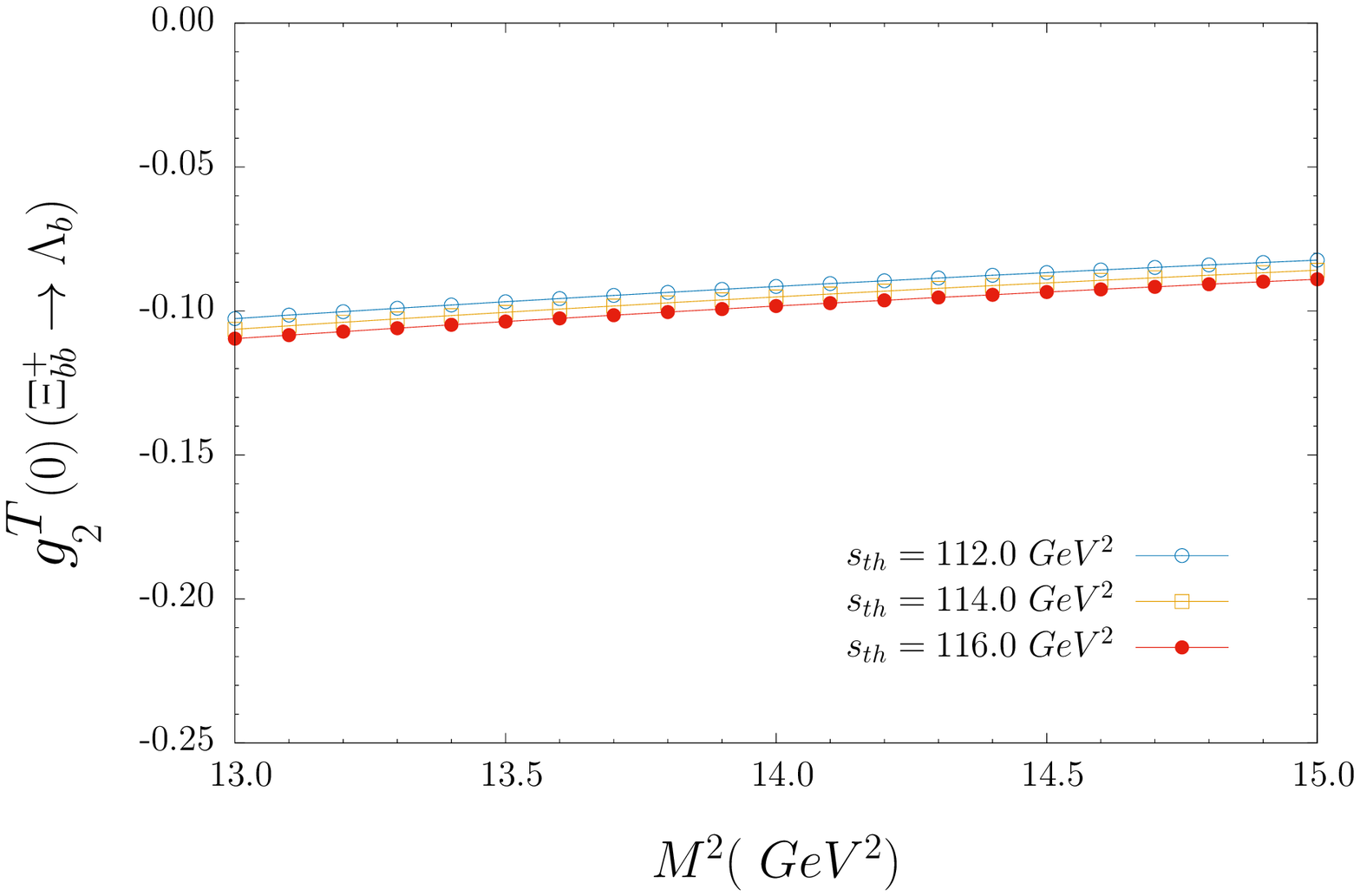}\\
\caption{
The variation of form factors (at $q^2=0$) with respect to the Borel mass parameter at the fixed values of $s_{th}$ is shown. 
}
\label{fig:fig1}
\end{figure}

The light cone sum rules are reliable in the region where $q^2$ is not too large, and our calculation indicates that LCSR predictions are reliable up to  $q^2 \lesssim 0.5~GeV^2$ for $\Lambda_c$ and $q^2 \lesssim 10~GeV^2$ for $\Lambda_b$ baryon, respectively. In order to study the stability of the form factors with respect to the variation of $M_B^2$ and $s_{th}$, we present the dependency of the form factors on Borel mass square  at the fixed values of $s_{th}$ in Figure~\ref{fig:fig1} for $\Xi_{bb} \to \Lambda_{b}$ transition at $q^2 =0$. Note that the choice of this point is due to the fact that LCSR works well at this point.  From this Figure, we see that all the form factors exhibit good stability with respect to the variation of $M_{B}^2$ in its working region. And we deduce the values of the form factors at $q^2=0$ presented in Table~\ref{tab:4}. Performing the similar analysis for the form factors responsible for $\Xi_{cc} \to \Lambda_{c}$ transition and at $q^2=0$ point, their  values are also presented in Table~\ref{tab:4}. 

To extend the obtained results of the form factors to the physical region $ 4 m_l^2 \leq q^2 \leq (M^2 - m_2)^2$, some parameterization is needed. For this goal, we look for a parameterization so that in the region where LCSR is reliable, the result of this parameterization and predictions of LCSR coincide.
\begin{table*}[hbt]
  \centering
  \renewcommand{\arraystretch}{1.4}
  \setlength{\tabcolsep}{7pt}
  \begin{tabular}{ccc}
    \toprule
Channels                                & $M_B^2(GeV^2)$ & $s_{th} (GeV^2)$ \\
    \midrule
$\Xi_{c c} \rightarrow \Lambda_{c} l l$ & $6 \pm 1  $    & $16 \pm 1 $       \\
$\Xi_{b b} \rightarrow \Lambda_{b} l l$ & $14 \pm 1 $    & $114 \pm 2 $      \\
\bottomrule
  \end{tabular}
  \caption{The working regions of Borel mass parameter $M_B^2$ and the continuum threshold $s_{th}$.}
  \label{tab:3}
\end{table*}
Numerical results show that the best parameterization of the form factors which satisfy the above criteria can be represented as a double pole parameterization
\begin{equation}
  \label{eq:9}
  F_i\left(q^{2}\right)=\frac{F_i(0)}{1-\frac{q^{2}}{m_{i}^{2}} + \alpha_i \left(\frac{q^{2}}{m_{i}^{2}}\right)^{2}}.
\end{equation}
The obtained fitting parameters  $\alpha_i$ and $m_i$ are collected in Table~\ref{tab:4}.
\begin{table*}[hbt]
  \centering
  \renewcommand{\arraystretch}{1.3}
  \setlength{\tabcolsep}{5pt}
  \begin{tabular}{lcccc}
    \toprule
Channels                                 & $F$       & $F(0)$                          & $\alpha$             & $m_i$                    \\
    \midrule
$\Xi_{c c} \rightarrow \Lambda_{c} l^+ l^- $ & $f_{1}^T$ & $-1.25 \pm 0.14$            & $83.95 \pm 6.93$     & $4.15 \pm 0.11$           \\
$\Xi_{c c} \rightarrow \Lambda_{c} l^+ l^-$  & $f_{2}^T$ & $0.13 \pm 0.01$           & $0.607 \pm 0.004$    & $1.1878 \pm 0.0008$       \\
$\Xi_{c c} \rightarrow \Lambda_{c} l^+ l^-$  & $g_{1}^T$ & $0.51 \pm 0.06 $ & $-0.0934 \pm 0.0004$ & $1.47389 \pm 0.00005$      \\
$\Xi_{c c} \rightarrow \Lambda_{c} l^+ l^-$  & $g_{2}^T$ & $0.13 \pm 0.01$            & $0.607 \pm 0.003$    & $1.1879 \pm 0.0009$       \\
\midrule
$\Xi_{b b} \rightarrow \Lambda_{b} l^+ l^-$  & $f_{1}^T$ & $0.28 \pm 0.03 $          & $0.786 \pm 0.005 $   & $3.76 \pm 0.003$          \\
$\Xi_{b b} \rightarrow \Lambda_{b} l^+ l^-$  & $f_{2}^T$ & $-0.09 \pm 0.01 $ & $1.891 \pm 0.002$    & $4.659 \pm 0.001$         \\
$\Xi_{b b} \rightarrow \Lambda_{b} l^+ l^-$  & $g_{1}^T$ & $0.01 \pm 0.001$             & $10.19 \pm 5.19$     & $2.10 \pm 0.29$              \\
$\Xi_{b b} \rightarrow \Lambda_{b} l^+ l^-$  & $g_{2}^T$ & $-0.09 \pm 0.01 $ & $1.891 \pm 0.002$    & $4.659 \pm 0.001$            \\
\bottomrule
  \end{tabular}
  \caption{The values of the fitting parameters  $F(0)$, $\alpha$ and $m$.}
  \label{tab:4}
\end{table*}
%%%%%%%%%%%%%%%%%%%

Having the form of the form factors, our final goal is the calculation of the branching ratios of $\Xi_{cc}^{++} \to \Lambda_c l^+ l^-$ and $\Xi_{bb}^{++} \to \Lambda_b l^+ l^-$ transition. After straightforward calculations of the decay width for $\Xi_{bb}^{++} \to \Lambda_b l^+ l^-$  we find for,
\begin{equation}
  \label{eq:10}
  \frac{d \Gamma}{ds} =  \frac{G_F^2 \alpha_{em}^2 M}{4096 \pi^5} |V_{tb} V_{td}^*|^2 v \sqrt{\lambda(1,r,s)} \bigg[ T_1(s) + \frac{1}{3} T_2(s) \bigg]
\end{equation}
where  $\alpha_{em}$ is the fine structure constant, $v = \sqrt{1 - 4 m_l^2/q^2}$ is the velocity of the lepton and $\lambda(1,r,s) = 1 + r^2 + s^2 - 2 r -2 s - 2 r s$, $r = \frac{m^2}{M^2}$, $s=\frac{q^2}{M^2}$. The expressions of the $T_1(s)$ and $T_2(s)$ are presented in Appendix~\ref{sec:appb}.

The differential decay width for $\Xi_{cc} \to \Lambda_c l^+ l^-$ transition can easily be obtained from Eq.~\eqref{eq:10} by following replacements,
\begin{equation}
  \label{eq:11}
  V_{tb} V_{td} \to \sum V_{ci} V_{ui}^*, \hspace{0.5cm}  m_{\Lambda_b} \to m_{\Lambda_c}, \hspace{0.5cm}   M_{\Xi_{bb}} \to M_{\Xi_{cc}}, \hspace{0.5cm} \tau_{\Xi_{cc}} \to \tau_{\Xi_{bb}} 
\end{equation}
Integrating over $s$ in the region $\frac{4 m_l^2}{M^2} \leq s \leq (1 - \sqrt{r })^2$ and using the lifetimes of $\Xi_{cc}^+$ and $\Xi_{bb}$ baryons $\tau_{\Xi_{cc}^+} = 45 \times 10^{-15}~s$~\cite{LHCb:2017iph} and $\tau_{\Xi_{bb}^+} = 370 \times 10^{-15}~s$~\cite{LHCb:2018pcs},
we get the results for the branching ratio presented in Table~\ref{tab:5}.
\begin{table*}[hbt]
    \centering
  \renewcommand{\arraystretch}{1.4}
  \setlength{\tabcolsep}{7pt}
\begin{tabular}{lccc}
\toprule
  Channels                                                      & Our result           &      \cite{Xing:2018lre}                     &  \cite{Hu:2020mxk}           \\
  \midrule
$\Xi_{c c}^{++} \rightarrow \Lambda_{c}^{+} l^{+} l^{-}$    & $(1.16 \times 10^{-13})$       &       ---                    &    ---       \\
  \midrule
$\Xi_{b b}^{++} \rightarrow \Lambda_{b}^{+} e^{+} e^{-}$        & $(1.14 \times 10^{-9})$    &    $3.63 \times 10^{-9}$  &    $2.33 \times 10^{-9}$          \\
$\Xi_{b b}^{++} \rightarrow \Lambda_{b}^{+} \mu^{+} \mu^{-}$    & $(4.04 \times 10^{-10})$   &   $3.55 \times 10^{-9}$    &   $2.24 \times 10^{-9}$            \\
$\Xi_{b b}^{++} \rightarrow \Lambda_{b}^{+} \tau^{+} \tau^{-}$  & $(4.38 \times 10^{-12})$   & $9.86 \times 10^{-10}$     &    $8.49 \times 10^{-11}$           \\
  \bottomrule
    \end{tabular}
  \caption{Branching ratios for the considered decays.}
  \label{tab:5}
\end{table*}

In this Table, we also present the results obtained in the light-front approach~\cite{Xing:2018lre,Hu:2020mxk}. From the comparison, we see that our  branching ratios' results are smaller than those presented in \cite{Xing:2018lre,Hu:2020mxk}. The main source of the discrepancy between the predictions of the different approaches is the values of the form factors. Our final remark to this section is that our result can be improved by taking into account $\mathcal{O}(\alpha_s)$ corrections to the distribution amplitudes of $\Lambda_Q$.

%%%%%%%%%%%%%%%%%%%%%%%%%
\section{Conclusion}
\label{conclusion}
In the present work, we calculate the form factors of the $\Xi_{cc}^+ \to \Lambda_c^+ l^+ l^-$ and $\Xi_{bb} \rightarrow \Lambda_b^0 l^+ l^-$ induced by tensor currents within the light cone sum rules. Using the obtained results for the form factors, we estimate the corresponding branching ratios induced by neutral currents $c \to u l^+ l^-$ and $b \to d l^+ l^-$ transitions. We also compared our results with the predictions obtained in the framework of the light-front approach and found out that our predictions on the corresponding branching ratios are smaller than the ones predicted in the light-front approach.
The predictions obtained for the branching ratio, especially for $\Xi_{bb} \rightarrow \Lambda_b^0 l^+ l^-$ hopefully, will be inspected at LHCb; however, the branching ratio for  $\Xi_{cc}^+ \to \Lambda_c^+ l^+ l^-$ is too small to be detected in the near future with current detector technologies.
\section*{Acknowledgements}
The authors thank Y.J.Shi for useful discussions.

%%%%%%%%%%%%%%%%%%%%%%%%%%%%%%%%%% 
 \bibliographystyle{utcaps_mod}
 \bibliography{../all.bib}

%%%%%%%%%%%%%%%%%%%%%%%%%%%%%%%%%%

\newpage
\appendix
\section{Explicit expressions of the spectral densities}
\label{sec:appa}
In this appendix, we present the explicit expressions of $\rho_n^{(i)}(u,w)$ entering to the Eq.~\eqref{eq:34}
\begin{equation}
  \label{eq:38}
  \begin{split}
\rho_1^{(1)}(u,w) &= 
- f^{(2)} w \psi_3^{(s)}(u,w)~,  \\
\rho_2^{(1)}(u,w) &=  
- \bar{u} \Big\{2 f^{(2)} \Big[\bar{u} w - (q\mcdot v) \Big]  
\widehat{\psi}_3^{(\sigma)}(u,w) +
f^{(1)} m_Q \Big[ \widehat{\psi}_4 (u,w) -  \widehat{\psi}_2 (u,w) \Big] \Big\}~, \\
\rho_1^{(2)}(u,w) &= 
- 2 f^{(2)} \bar{u} \Big[w^2 \psi_3^{(s)}(u,w) +  
\widehat{\psi}_3^{(\sigma)}(u,w)\Big]~,   \\
\rho_2^{(2)}(u,w) &= 
4 f^{(2)} \bar{u}^2 w (q\mcdot v)  
\widehat{\psi}_3^{(\sigma)}(u,w)~,  \\
\rho_1^{(3)}(u,w) &= 
f^{(2)} w \psi_3^{(s)}(u,w)~,   \\  
\rho_2^{(3)}(u,w) &= 
\bar{u} \Big\{2 f^{(2)} \Big[\bar{u} w - (q\mcdot v)\Big]
\widehat{\psi}_3^{(\sigma)}(u,w) -
f^{(1)} m_Q \Big[ \widehat{\psi}_4 (u,w) -  \widehat{\psi}_2 (u,w)\Big] \Big\}~,\\
\rho_1^{(4)}(u,w)  &= 
- 2 f^{(2)} \bar{u} \Big[w^2 \psi_3^{(s)}(u,w) +
\widehat{\psi}_3^{(\sigma)}(u,w) \Big],  \\  
\rho_2^{(4)}(u,w) &= - 4 f^{(2)} \bar{u}^2 w (q\mcdot v)
\widehat{\psi}_3^{(\sigma)}(u,w)~. 
\end{split}
\end{equation}
where $\bar{u} = 1-u$. 
The functions $\widehat{\psi}$ entering the above equation is defined as
\begin{equation}
  \label{eq:20}
  \begin{split}
  \hat{\psi}(u, \omega) &= \int_0^{\omega} dy y \psi(u,y)~.
\end{split}
\end{equation}
\newpage
\section{The differential width for $\Xi_{QQ} \to \Lambda_Q l^+ l^-$ decay}
\label{sec:appb}
Here, we present the differential width for $\Xi_{bb} \to \Lambda_b l^+ l^-$ decay for completeness (see~\cite{Aliev:2018hyy}). As we noted in the main body of the text, the differential decay width for this transition is given by,
\begin{equation}
  \label{eq:100}
  \frac{d \Gamma}{ds} = \frac{G_F^2 \alpha_{em}^2 M}{4096 \pi^5} |V_{tb} V_{td}^*|^2 v \sqrt{\lambda(1,r,s)} \bigg[ T_1(s) + \frac{1}{3} T_2(s) \bigg]
\end{equation}
The functions $T_1(s)$ and $T_{2}(s)$ are 
\begin{equation}
\begin{array}{l}
T_{1}(s)=8 M^{2}\left\{(1-2 \sqrt{r}+r-s)\left[4 m_{\ell}^{2}+M^{2}(1+2 \sqrt{r}+r+s)\right]\left|F_{1}\right|^{2}\right.\\
-\left[4 m_{\ell}^{2}(1-6 \sqrt{r}+r-s)-M^{2}\left((1-r)^{2}-4 \sqrt{r} s-s^{2}\right)\right]\left|F_{4}\right|^{2}\\
+(1-2 \sqrt{r}+r-s)\left[4 m_{\ell}^{2}(1+\sqrt{r})^{2}+M^{2} s(1+2 \sqrt{r}+r+s)\right]\left|F_{2}\right|^{2}\\
+M^{2} s\left[(-1+r)^{2}-4 \sqrt{r} s-s^{2}\right] v^{2}\left|F_{4}\right|^{2}\\
+4 m_{\ell}^{2}(1+2 \sqrt{r}+r-s) s\left|F_{6}\right|^{2}\\
+(1+2 \sqrt{r}+r-s)\left[4 m_{\ell}^{2}+M^{2}(1-2 \sqrt{r}+r+s)\right]\left|G_{1}\right|^{2}\\
-\left[4 m_{\ell}^{2}(1+6 \sqrt{r}+r-s)-M^{2}\left((1-r)^{2}+4 \sqrt{r} s-s^{2}\right)\right]\left|G_{4}\right|^{2}\\
+(1+2 \sqrt{r}+r-s)\left[4 m_{\ell}^{2}(1-\sqrt{r})^{2}+M^{2} s(1-2 \sqrt{r}+r+s)\right]\left|G_{2}\right|^{2}\\
+M^{2} s\left[(1-r)^{2}+4 \sqrt{r} s-s^{2}\right] v^{2}\left|G_{5}\right|^{2}\\
+4 m_{\ell}^{2}(1-2 \sqrt{r}+r-s) s\left|G_{6}\right|^{2}\\
-4(1+\sqrt{r})(1-2 \sqrt{r}+r-s)\left(2 m_{\ell}^{2}+M^{2} s\right) \operatorname{Re}\left[F_{1}^{*} F_{2}\right]\\
-4 M^{2}(1+\sqrt{r})(1-2 \sqrt{r}+r-s) s v^{2} \operatorname{Re}\left[F_{4}^{*} F_{5}\right]\\
-8 m_{\ell}^{2}(1-\sqrt{r})(1+2 \sqrt{r}+r-s) \operatorname{Re}\left[F_{4}^{*} F_{6}\right]\\
-4(1-\sqrt{r})(1+2 \sqrt{r}+r-s)\left(2 m_{\ell}^{2}+M^{2} s\right) \operatorname{Re}\left[G_{1}^{*} G_{2}\right]\\
\left.-4 M^{2}(1-\sqrt{r})(1+2 \sqrt{r}+r-s) s v^{2} \operatorname{Re}\left[G_{4}^{*} G_{5}\right]\right]\\
\left.+8 m_{\ell}^{2}(1+\sqrt{r})(1-2 \sqrt{r}+r-s) \operatorname{Re}\left[G_{4}^{*} G_{6}\right]\right\} \text {, }
\end{array}
\end{equation}
\begin{equation}
T_{2}(s)=-8 M^{4} v^{2} \lambda(1, r, s)\left[\left|F_{1}\right|^{2}+\left|F_{4}\right|^{2}+\left|G_{1}\right|^{2}+\left|G_{4}\right|^{2}-s\left(\left|F_{2}\right|^{2}+\left|F_{5}\right|^{2}+\left|G_{2}\right|^{2}+\left|G_{5}\right|^{2}\right)\right]
\end{equation}
where,
\begin{align}
F_{1} &=C_{9} f_{1}-\frac{2 m_{b}}{M} C_{7} f_{1}^{T} \nonumber \\
F_{2} &=C_{9} f_{2}+\frac{2 m_{b}}{q^{2}} M f_{2}^{T} \nonumber \\
F_{3} &=C_{9} f_{3}-\frac{2 m_{b}}{q^{2}} C_{7}\left(M-m\right) f_{1}^{T} \nonumber \\
G_{1} &=C_{9} g_{1}-\frac{2 m_{b}}{M} C_{7} g_{1}^{T} \nonumber \\
G_{2} &=C_{9} g_{2}+\frac{2 m_{b}}{q^{2}} M g_{2}^{T} \nonumber \\
G_{3}&=C_{9} g_{3}-\frac{2 m_{b}}{q^{2}} C_{7}\left(M+m \right) g_{1}^{T} \nonumber \\
F_{4}&=C_{10} f_{1} \nonumber \\
F_{5}&=C_{10} f_{2} \nonumber \\
F_{6}&=C_{10} f_{3} \nonumber \\
G_{4}&=C_{10} g_{1} \nonumber \\
G_{5}&=C_{10} g_{2} \nonumber \\
G_{6}&=C_{10} g_{3} 
\end{align}
The differential width for the $\Xi_{cc} \to \Lambda_{c} l^+ l^-$ can be obtained from the result presented above by replacements given by Eq.~\eqref{eq:11}.

%%%%%%%%%%%%%%%%%%%%%%%%%%%%%%%%%%

\end{document}